\documentclass[twocolumn]{aastex631}

\begin{document}

\title{Left Ringing: Betelgeuse Illuminates the Connection Between Convective outbursts, Mode switching, and Mass Ejection in Red Supergiants}
\correspondingauthor{Morgan MacLeod}
\email{morgan.macleod@cfa.harvard.edu}

\author[0000-0002-1417-8024]{Morgan MacLeod}
\affiliation{Center for Astrophysics $\vert$ Harvard $\&$ Smithsonian 
60 Garden Street, MS-16, Cambridge, MA 02138, USA}
\affiliation{Institute for Theory and Computation}

\author[0000-0003-3062-4773]{Andrea Antoni}
\affiliation{Astronomy Department and Theoretical Astrophysics Center, University of California, Berkeley, CA 94720, USA}

\author[0000-0001-6169-8586]{Caroline D. Huang}
\affiliation{Center for Astrophysics $\vert$ Harvard $\&$ Smithsonian 
60 Garden Street, MS-16, Cambridge, MA 02138, USA}

\author[0000-0002-8985-8489]{Andrea Dupree}
\affiliation{Center for Astrophysics $\vert$ Harvard $\&$ Smithsonian 
60 Garden Street, MS-16, Cambridge, MA 02138, USA}

\author[0000-0003-4330-287X]{Abraham Loeb}
\affiliation{Center for Astrophysics $\vert$ Harvard $\&$ Smithsonian 
60 Garden Street, MS-16, Cambridge, MA 02138, USA}
\affiliation{Institute for Theory and Computation}

\begin{abstract}
Betelgeuse, the nearest red supergiant, dimmed to an unprecedented level in early 2020. The star emerged from this Great Dimming episode with its typical, roughly 400-day pulsation cycle halved, and a new dominant period of around 200 days. The dimming event has been attributed to a surface mass ejection, in which rising material drove shocks through the stellar atmosphere and expelled some material, partially obscuring the star as it formed molecules and dust. In this paper, we use hydrodynamic simulations to reveal the connections between Betelgeuse’s vigorously convective envelope, the surface mass ejection, and the pulsation mode switching that ensued. An anomalously hot convective plume, generated rarely but naturally in the star’s turbulent envelope, can rise and break free from the surface, powering an upwelling that becomes the surface mass ejection. The rising plume also breaks the phase coherence of the star’s pulsation, causing the surface to keep expanding even as the deeper layers contract. This drives a switch from the 400-day fundamental mode of pulsation, in which the whole star expands and contracts synchronously, to the 200-day first overtone, where a radial node separates the interior and exterior of the envelope moving in opposite phase. We predict that the star’s convective motions will damp the overtone oscillation and Betelgeuse will return to its previous, 400-day fundamental mode pulsation in the next 5-10 years. With its resolved surface and unprecedentedly detailed characterization, Betelgeuse opens a window to episodic surface mass ejection in the late-stage evolution of massive stars.
\end{abstract}


\section{Introduction} \label{sec:intro}

Betelgeuse ($\alpha$ Orionis) is a bright, nearby supergiant star. Its angular size of roughly 42~mas is so large that it has been spatially resolved in Hubble Space Telescope (HST) ultraviolet imaging \citep{1996ApJ...463L..29G}, interferometric imaging \citep[e.g.][]{2009A&A...508..923H,2018A&A...609A..67K,2021Natur.594..365M}, and Hubble Space Telescope (HST) spectroscopy \citep[e.g.][]{2000ApJ...545..454L,2020ApJ...899...68D}. Betelgeuse's distance, and thus radius and mass, remain uncertain, but the star is likely $16-20M_\odot$ and $\sim 700-900R_\odot$, implying a distance of $\sim 170$~pc \citep{2020ApJ...902...63J}, although other measures suggest a distance of $\sim 220$~pc \citep{2017AJ....154...11H}. Betelgeuse's effective temperature is $\sim 3600$~K \citep{2020ApJ...891L..37L}, and the star's luminosity is $\sim 10^5L_\odot$. Betelgeuse is likely burning helium in its core \citep{2020ApJ...891L..37L} and is vigorously convecting in its envelope \citep[e.g.][]{2021Natur.594..365M}.

Betelgeuse has long been known to be variable, exhibiting semi-irregular pulsations of $\sim 400$~d period and $\sim 0.5$~mag photometric amplitude. This optical variability has been associated with a radial pulsation in the star's fundamental mode driven by the $\kappa$ (opacity) mechanism \citep{2020ApJ...902...63J}. However, in late 2019 into early 2020 Betelgeuse underwent a historic dimming, fading by nearly a magnitude below its typical range. Observational studies have revealed that this historic fading was associated with mass ejection from the stellar surface \citep{2020ApJ...899...68D,2022ApJ...936...18D}. Shocks were observed tracing outward through the atmosphere, leaving a cooler photosphere in their wake \citep{2021A&A...650L..17K}.  Though a number of initial proposals were raised, the eventual dimming appears to be attributable to dust formation in ejected material partially obscuring the star \citep{2020ApJ...899...68D,2021Natur.594..365M,2022ApJ...936...18D}. 

Since the dimming, Betelgeuse's light and radial velocity curves have  been markedly different from its past. There has been a steady return toward the star's previous brightness, likely associated with the expansion and clearing of dusty ejecta. However, the variability in the past three years has been more rapid, with $\lesssim 200$~d cycles typical, as opposed to the previous $\sim$400~d variability. 

In this paper, we propose a model that unites the surface mass ejection and the subsequent behavior. In our proposed scenario, an unusually large hot blob arises within the turbulent stellar interior.  As this buoyant blob of fluid rises, its leading edge drives a shock through the stellar atmosphere. This breakout expells some material that goes on to form the dusty ejecta. After its initial breakout, the plume persists for some time, interfering with ongoing pulsation of star (which has a typical magnitude, $\sim 1$~km~s$^{-1}$, which is significantly less than the $\sim 10$~km~s$^{-1}$ convective velocity), and causing a switch from the fundamental mode to higher-frequency overtones.  Over time, the same convection that gave rise to the outburst may damp these overtone oscillations, suggesting that Betelgeuse will return to its previous, primarily fundamental-mode pulsation in the coming years. 

The remainder of this paper is organized as follows.  Section 2 describes our methods in creating a numerical model of a convective outburst. Section 3 analyzes this representative model through phases of outburst, oscillation, and damping. Section 4 reviews the observational evidence surrounding Betelgeuse's Great Dimming and discusses how our modeling provides some interpretive framework for these data. In Section 5, we conclude.

%
%

\section{Numerical Models}\label{sec:method}
\subsection{Stellar Evolution Model}
We construct an approximate model of Betelgeuse using the stellar evolution code MESA \citep{2011ApJS..192....3P,2013ApJS..208....4P,2015ApJS..220...15P,2018ApJS..234...34P,2019ApJS..243...10P}. We follow the basic method outlined by \citet{2020ApJ...902...63J}, evolving a non-rotating star of Betelgeuse's approximate mass until its radial oscillation period and surface temperature match the observational constraints of $\sim 400$~d and $\sim 3600$~K, respectively \citep{2020ApJ...902...63J,2020ApJ...891L..37L}. 

Based on the results of \citet{2020ApJ...902...63J}, we adopt an initial (zero-age main sequence) mass of $18M_\odot$, and an initial metal fraction of $Z=0.02$. We apply MESA's ``Dutch" wind scheme with a coefficient of 1.0. We apply the Ledoux mixing criterion, a mixing-length $\alpha_{\rm mlt}=2.0$, thermohaline coefficient of 2.0, and semiconvection coefficient of $\alpha_{\rm semi}=0.01$. We adopt spatial and temporal resolution factors {\tt mesh\_delta\_coeff}=1 and {\tt time\_delta\_coeff}=1. 

We evolve this model forward through hydrogen burning and into core helium burning, until the mode frequency (evaluated as described in the following subsection) approximates the observed value. Our eventual model has an age of 8.6~Myr, an effective temperature of 3569~K, and a luminosity of $L=8.69\times10^4L_\odot$. The photosphere radius is $R=770.6R_\odot$. The model's current mass is $M=17.58M_\odot$, of which $M_{\rm env}=13.34M_\odot$ makes up the  hydrogen-rich envelope, and $4.24M_\odot$ is in the helium core. The model is relatively early in its core helium burning evolution, such that the central helium abundance is still 96.4\%.

\subsection{Oscillation Calculation}
We compute the characteristic oscillation frequencies of our Betelgeuse model using the oscillation code GYRE \citep{2013MNRAS.435.3406T,2018MNRAS.475..879T,2020ApJ...899..116G,2023arXiv230106599S}. To do so, we save output files from MESA in GYRE-compatible format in closely-spaced intervals after it develops an extended, convective envelope. To understand characteristic frequencies of the envelope, we excise the core, applying a zero-displacement boundary condition near the base of the convective envelope at $r/R=0.05$.  We then compute the nonadiabatic frequency of the radial fundamental mode  ($l=m=0$, $n_{\rm pg}=1$). Our selected model has a dimensionless frequency of $\omega_0 = 1.46 (GM/R^3)^{1/2}$, or a mode period, $2\pi/\omega_0 = 406$~d. 

To understand the response of the envelope to a range of modes, we compute the frequencies of the first several overtones of the radial and low-degree nonradial modes of our model. We consider $l=0-4$, $n_{\rm pg}=0-3$ modes, those with zero to three radial inflections in the convective envelope. We ignore Betelgeuse's slow rotation, which means that there is no prograde versus retrograde frequency splitting in our model. We summarize these frequencies in Table \ref{tab:modes}, and show some representative eigenfunctions in Figure \ref{fig:ef}. We also list the mode dimensionless growth rate, $\eta$, where $\eta>0$ indicates unstable, growing modes \citep{2013MNRAS.435.3406T,2018MNRAS.475..879T,2020ApJ...899..116G}.

\subsection{Hydrodynamic Simulation}

To qualitatively explore the process of a convective plume bursting through the stellar surface we perform a hydrodynamic simulation. Our model is based on {\tt Athena++} \citep{2020ApJS..249....4S} and the methodology of \citet{2022MNRAS.511..176A,2023arXiv230105237A} to create a convective model in quasi-hydrostatic equilibrium. \citet{2022MNRAS.511..176A} describe and validate the approach in detail. Here, we summarize some key features of these hydrodynamic models and focus on their application to our scenario. 

The baseline model of \citet{2022MNRAS.511..176A} is a hydrostatic, non-rotating, convecting stellar envelope. The model is computed in dimensionless length, mass, and time scales, as described in detail by \citet{2022MNRAS.511..176A}. The self-gravity of the envelope is ignored, and an ideal-gas equation of state with $\gamma_{\rm ad}=1.4762$ is applied. In addition, heat is added at the envelope base, and removed near its limb. The heating and cooling terms are imposed to drive  convective turnover, but not necessarily to mimic the exact structure of, for example, the stellar photosphere \citep[see equations 16--18 of][]{2022MNRAS.511..176A}.  Since we are interested in the appearance of phenomena near the photosphere, we changed the cooling prescription by varying the characteristic timescale over which cooling is applied \citep[equation 18 of][]{2022MNRAS.511..176A} outside of the cooling radius. Our fiducial choice is $10$ times the local dynamical time, which is different from \citet{2022MNRAS.511..176A}'s fiducial choice of $10^{-2}$ local dynamical times. Starting from models relaxed into thermal equilibrium by \citet{2022MNRAS.511..176A}, we let our models adjust to a new hydrostatic equilibrium with the modified cooling over 1000 simulation time units (or 44 surface dynamical times). We ignore Betelgeuse's $\sim 30$~yr rotation period since it is much longer than our simulation duration.

The simulation volume is $(40r_0)^3$ where $r_0$ is the code unit of length (and we treat the surface as being at $8r_0$). This volume is covered by a $256^3$ base mesh, with 6 nested levels of $256^3$ mesh  covering each factor of two smaller in distance from the origin. 
We rescale these simulations in what follows to a total mass of $\sim 17.58 M_\odot$ and radius at $8r_0$ of $770.6R_\odot$, meant to be broadly representative of Betelgeuse -- see Section 3.6 of \citet{2022MNRAS.511..176A} for a more detailed description of the simulation units. In these physical units, the model's luminosity is  $2.3\times10^5 L_\odot$, or about twice that of Betelgeuse. 

\begin{figure}[tbp]
    \centering
    \includegraphics[width=\columnwidth]{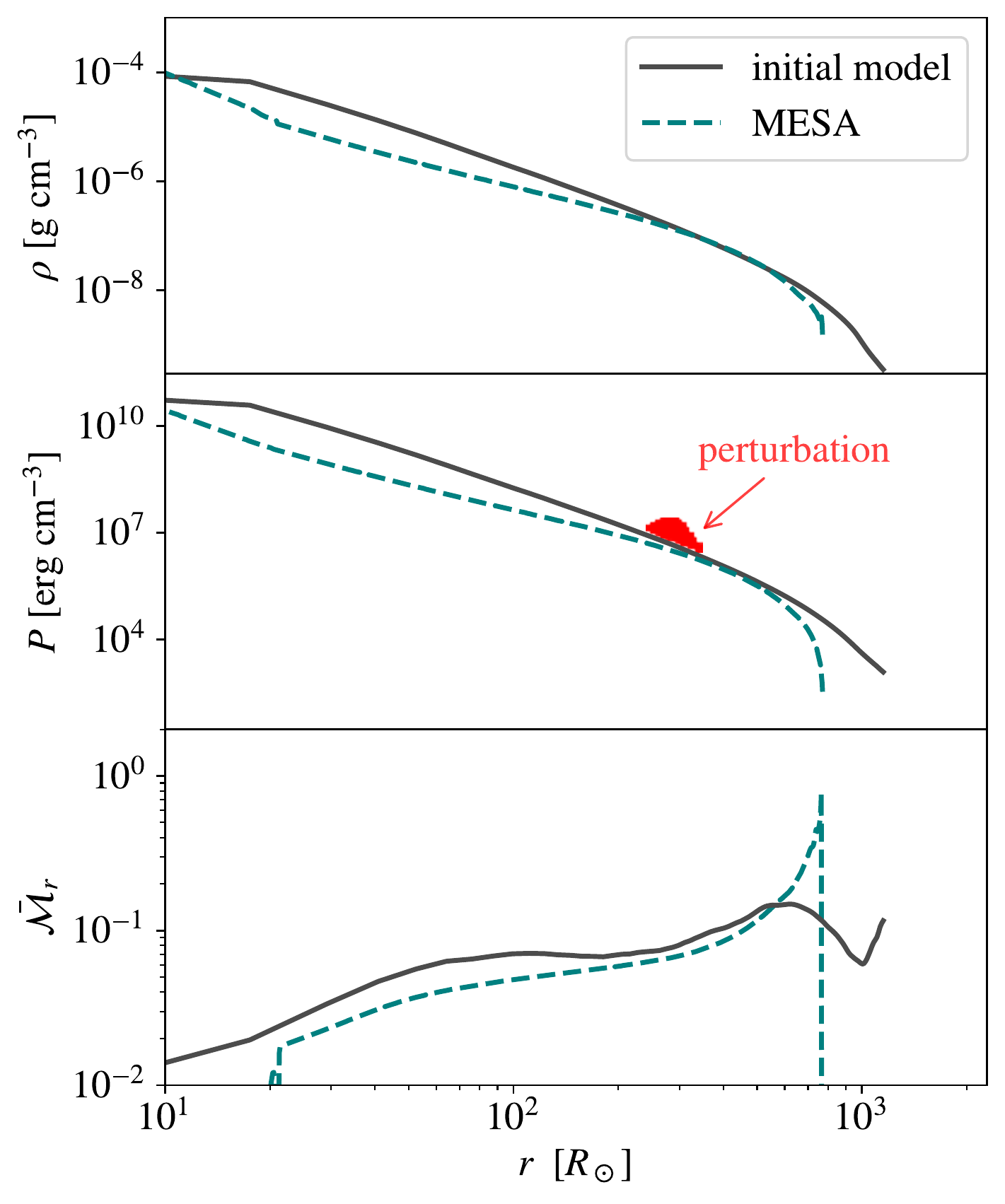}
    \caption{Radial structure of a MESA 1D stellar evolution model approximating Betelgeuse, where we plot the mass density, pressure, and radial convective Mach number in the three panels. We compare to spherically-averaged quantities from the initial state of our hydrodynamic simulation, before a pressure perturbation is added. The pressure perturbation itself is shown in red, with the values of individual zones plotted directly.  }
    \label{fig:ic}
\end{figure}

In Figure \ref{fig:ic} we compare the initial hydrodynamic model to our MESA stellar evolution model. The structure overall broadly mimics that MESA model in the bulk of the envelope and departs slightly near the limb, with shallower gradients of pressure and density -- and resulting lower radial Mach numbers -- out of numerical necessity. 
Other basic characteristics of the model compare favorably to what is known about convecting red supergiants. Large-scale eddies carry most of the convective flux, with a cascade of features to smaller scales clearly seen in \citet{2022MNRAS.511..176A}'s Figure 4. The spherically-averaged convective Mach number ranges from $\sim 0.01 - 0.2$, from the deep envelope to the surface. The large convective Mach number near the surface leads directly to the sorts of large scale flows and instabilities observed in Betelgeuse's distorted and actively convecting photosphere.

To model the emergence of a particularly hot, buoyant fluid parcel, we manually raise the pressure in an off-center region in the initial model, then allow the system to freely evolve. We apply this enhancement as a multiplicative factor to the pre-existing pressure over a Gaussian region. This is shown over the spherically-averaged pressure profile in Figure \ref{fig:ic}.  We experimented with the size and magnitude of this enhancement. Our fiducial model has a peak factor of 4.0 and width of $\sigma= 24R_\odot$ applied off center at a distance of half the stellar radius, $335R_\odot$,  in the $+x$-direction. The thermal energy added is $\approx 4\times10^{-4}$ of the thermal energy of the entire envelope, or $2.7\times 10^{45}$~erg.

We note that significantly more sophisticated radiation hydrodynamic models exist for convecting supergiant stars \citep[e.g.][]{2011A&A...535A..22C,2012JCoPh.231..919F,2017A&A...600A.137F,2019A&A...623A.158H,2023A&A...669A..49A,2022ApJ...929..156G}. Indeed, these models self-consistently yield the development of turbulent convection and the boiling, time-variable stellar surface that results. While our approach intends to reveal the basic physics, future study of Betelgeuse's outburst in this sort of framework would be very productive and would allow more quantitative model-to-data comparisons. For example, because the thermodynamics of our model system's surface are crudely approximated, we cannot compare to observed photospheric temperature variations during the outburst.  Additionally, we note that studying the interaction of particularly strong convective plumes with pre-existing pulsations would be an extremely relevant extension of our current study \citep[e.g.][]{2023A&A...669A..49A}.

\section{Simulated Convective Outburst}\label{sec:results}
\subsection{Outburst}

\begin{figure*}[tbp]
    \centering
    \includegraphics[width=\textwidth]{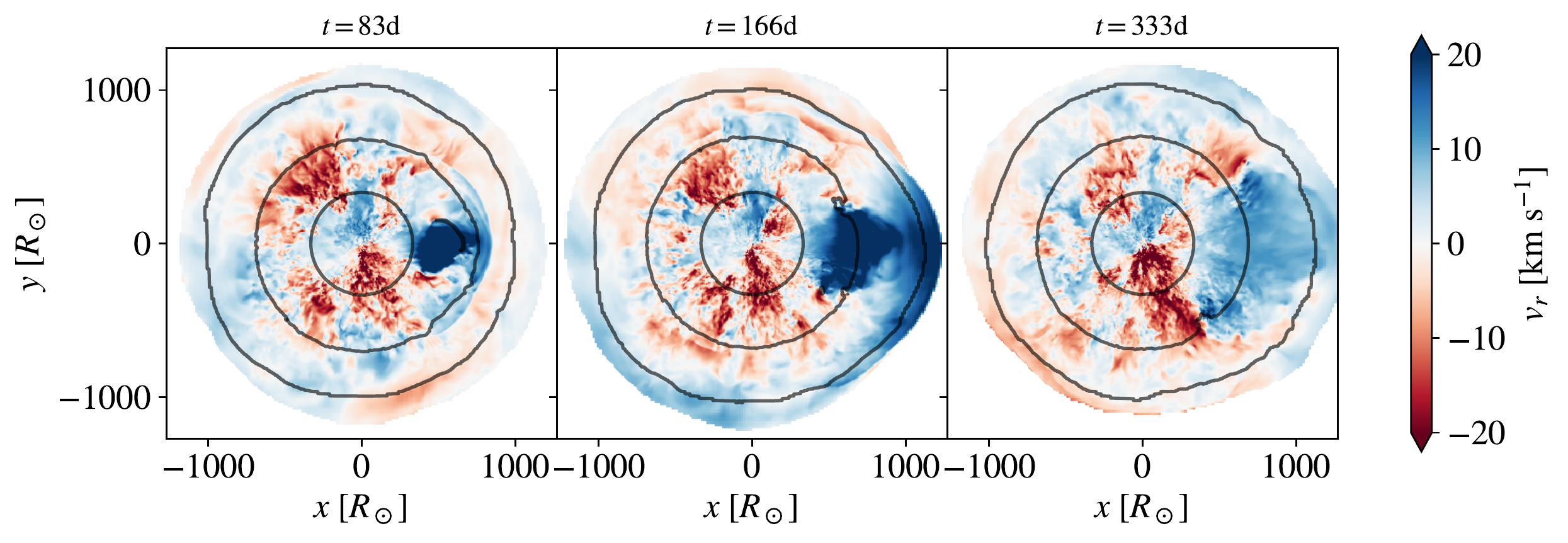}
 \includegraphics[width=0.85\textwidth]{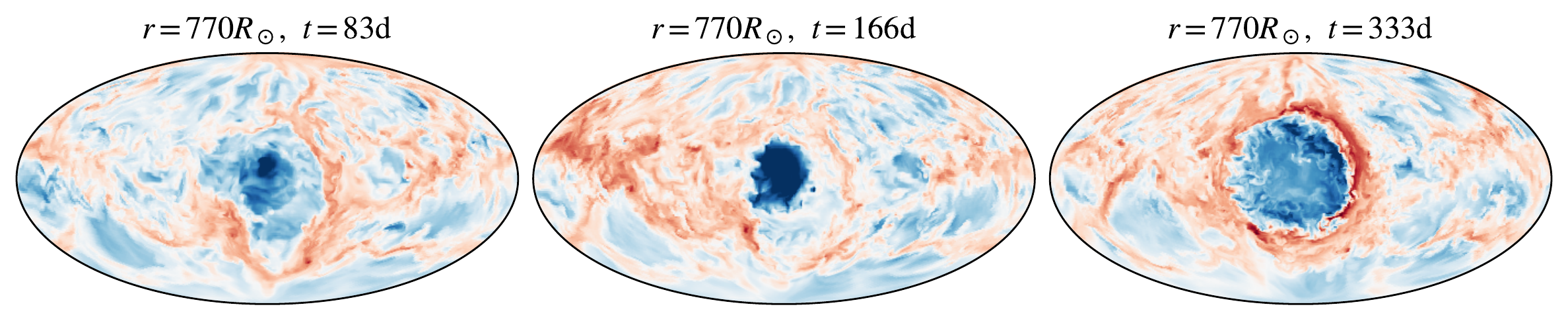}
    \caption{Slices through the computational domain of radial velocity relative to the stellar center as the hot, convective plume rises. Positive velocities indicate outflow, while negative velocities inflow in this frame of reference. The upper panels slice through the stellar envelope in the $x-y$ plane. The lower panels show slices at a radius of 770$R_\odot$ at the same times. The hot fluid, initially located at $(x,y)=(+335R_\odot, 0)$, begins rising and spreading over several hundred days as it breaks out of the stellar envelope.  }
    \label{fig:slice}
\end{figure*}

\begin{figure}
    \centering
    \includegraphics[width=\columnwidth]{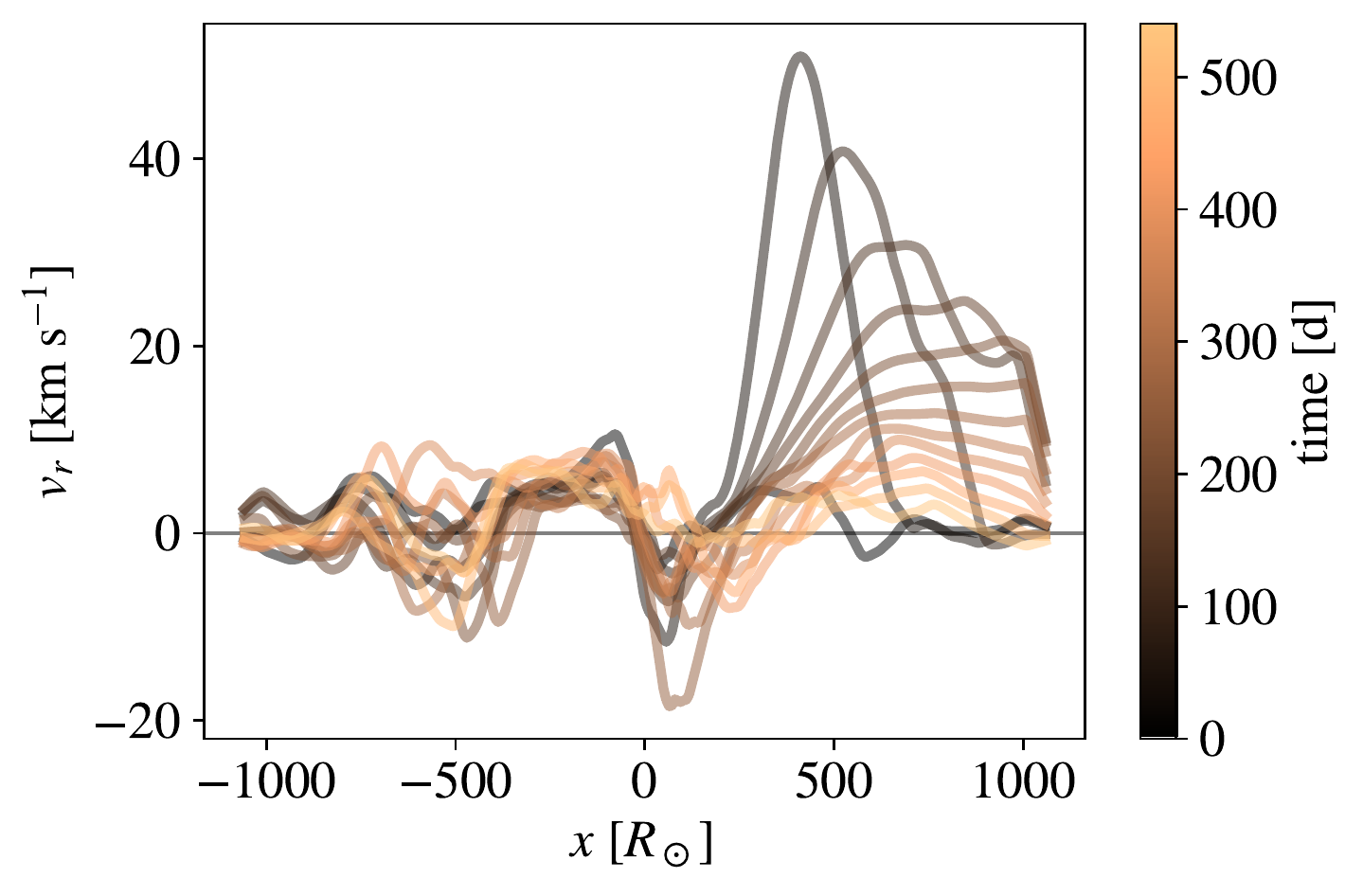}
    \caption{ Radial velocities relative to the stellar center along a ray located along the $x$-axis of the computational domain. For readability, velocities are smoothed with a 100$R_\odot$ moving top-hat kernel. In this time series, the outflowing motion of the hot convective plume decreases in intensity as the plume spreads and begins to break out.   }
    \label{fig:vr_ray}
\end{figure}

We hypothesize that the process that drives this outburst begins with a random collision in the turbulent flow. When two fluid parcels collide, their ram pressure $P_{\rm ram} \sim \rho v^2$ is converted into internal energy. The magnitude of a typical convective pressure perturbation is thus $\delta P/P \sim \rho v^2 / P$. For an ideal gas law, $c_s^2 = \gamma_{\rm ad} P/\rho$, so this becomes $\delta P/P \sim \gamma_{\rm ad}  \mathcal{M}^2$, where $\mathcal{M}=v/c_s$ is the convective Mach number. The spherically averaged radial component of the convective Mach number, shown in Figure \ref{fig:ic} is typically significantly less than unity. As a result, on average $\delta P/P \ll 1$. However, these are the averaged Mach numbers. What our models, and other models of turbulent convection, show is that the relevant velocity distributions are broad. Though the root-mean-square convective velocity is typically characterized by $\mathcal{M}\ll 1$, there  fluid parcels with particularly positive and negative buoyancy that obtain $\mathcal{M}\sim 1$ at all radii in our snapshots. If these parcels happen to collide, that can lead to an order-unity perturbation of the local pressure $\delta P/P \sim 1$. A particularly clear representation of this behavior is presented by \citet{2022ApJ...929..156G} who characterize convection in their radiation-hydrodynamic models of red supergiants. \citet{2022ApJ...929..156G}'s Figure 10 shows typical density fluctuations relative to the local mean.  The occurrence of these turbulent collisions is coincidental, as is the total mass of fluid involved. Though small fluid parcels collide and thermalize continuously, a large scale disturbance, like the one we postulate here, would be more rare.

We trace the emergence of a hot fluid blob from the convective envelope in Figure \ref{fig:slice}. As described in Section \ref{sec:method} and shown in Figure \ref{fig:ic}, our fiducial perturbation has a maximum amplitude of $\delta P/P =4$ and a Gaussian profile. It is imposed at half the envelope radius and has a Gaussian width of 3\% of the radius. The slices Figure \ref{fig:slice} show radial velocities in the $x-y$ plane within the convecting envelope. The hot parcel begins to rise rapidly, and drives a bow shock ahead of its motion. After this shock bursts free of the surface (beginning around the $t=83$~d panel), the bulk of the rising plume follows behind. 
Figure \ref{fig:vr_ray} plots radial velocities of material along the $y=0$ ray along the $x$-axis. Local velocities reach a maximum of $\sim 40$~km~s$^{-1}$ early in the breakout. As time proceeds, the rising plume expands in width and covers a larger surface area of the star, the characteristic outflow velocities gradually decrease.

Strikingly, the characterstic feature of a rising convective plume is that the radial velocities remain outflowing for several hundred days. This happens because the initially-spherical perturbation is radially stretched as it rises emerges over its characteristic buoyancy rise time. We note that the characteristic velocities we see at the stellar surface are lower than the $93$~km~s$^{-1}$ surface escape velocity. Thus, most of the rising plume is bound to the star. However, were our models better able to resolve the steep gradients of the stellar surface it is likely that a small fraction of the plume mass would escape  \citep{2021MNRAS.501.4266L}, or at least be raised to a few times the stellar radius where molecules and dust condense and couple the circumstellar material to the stellar radiation field. 

We performed additional experiments with different amplitude and magnitude perturbations. We found that even small amplitude perturbations (e.g. $\delta P/P=0.05$) deep in the envelope drive significant, fast shocks ahead of their rise. These shocks also become more spherical, covering at least a hemisphere of the stellar model. This is in part because of the higher energy density of the deeper layers, and in part because these shocks accelerate as they move through the steep portions of the outer envelope \citep{2021MNRAS.501.4266L}. By contrast,  shallower perturbations remain more local. At a given depth, a smaller amplitude of perturbation yields slower buoyant rise and lower surface velocities.

\subsection{Surface Signature of Outburst}

\begin{figure}
    \centering
\includegraphics[width=0.48\textwidth]{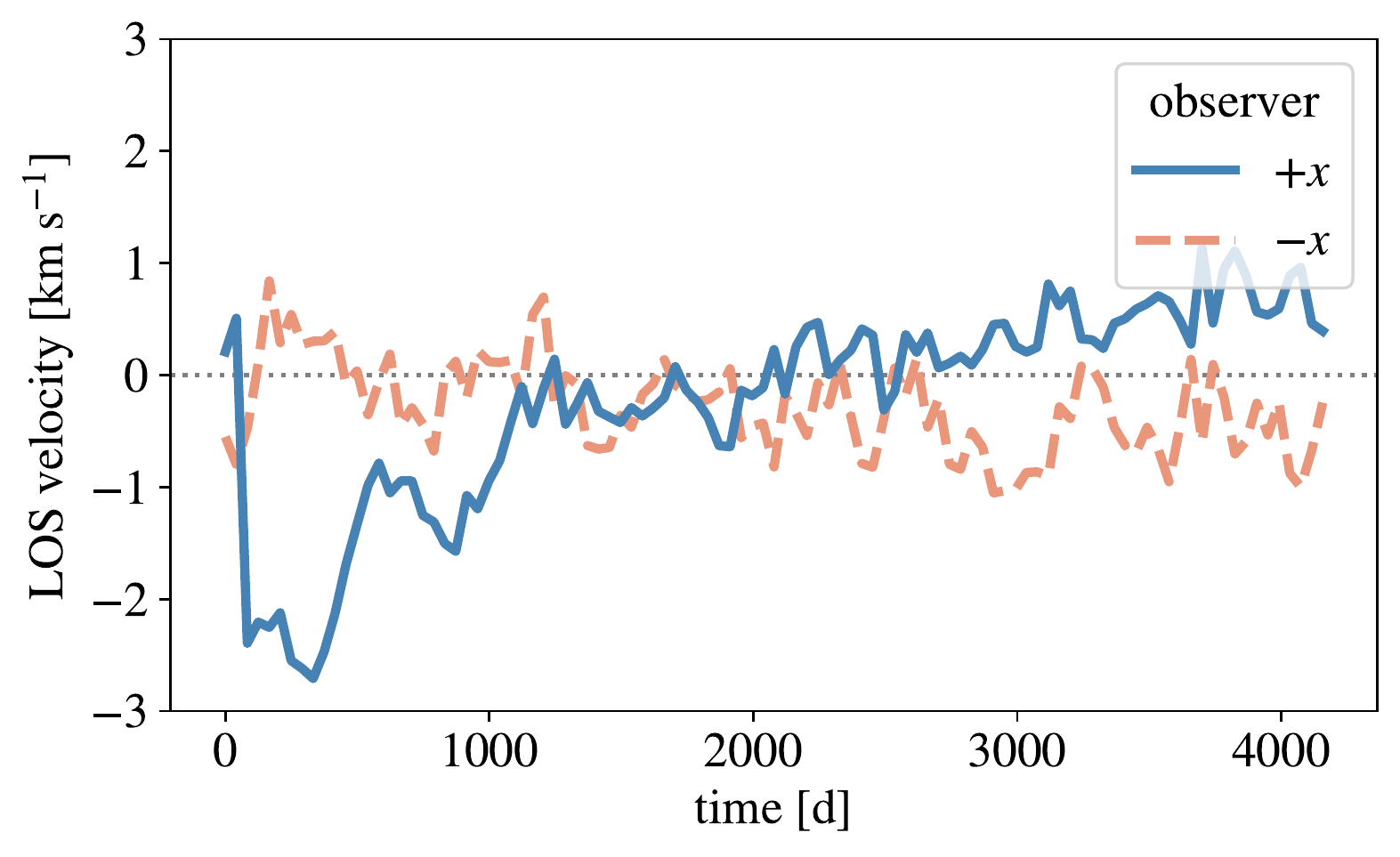}
    \caption{Line of sight velocity post-processed from the surface of our simulation snapshots at $r=770R_\odot$. Here we imagine observers in the $\pm x$-directions, and compute the disk-averaged line of sight velocity, where negative values imply outflow from the star toward the observer. An observer facing the breakout of the plume ($+ x$-direction) sees an extended period of negative velocities. By contrast an observer in the opposite direction sees little effect. Variability due to the evolving properties of convective plumes is exhibited from both viewing angles. }
    \label{fig:vlos}
\end{figure}

Next we consider the observable implication of the surface velocity distributions shown in Figure \ref{fig:slice}. Some spectra of Betelgeuse, for example with HST, are spatially resolved \cite[e.g.][]{2000ApJ...545..454L,2020ApJ...899...68D}, but ground based spectra average over the entire visible disk. To facilitate direct comparison to this observable, we post-process the surface flow in our model to compute the line-of-sight velocity for observers oriented in two perspectives in Figure \ref{fig:vlos}.  The observer in the $+x$-direction sees sustained negative line of sight velocities of $-2$-3~km~s$^{-1}$ (outflow) for roughly 400~d, before the line of sight velocity begins to shift rapidly back toward zero. The traces of the plume fully disappear after $\sim 1000$~d. By comparison an observer on the opposite side of the star, in the $-x$-direction, sees little to no signature of the convective breakout.  

\subsection{Dispersion}

\begin{figure*}[tbp]
    \centering
    \includegraphics[width=\columnwidth]{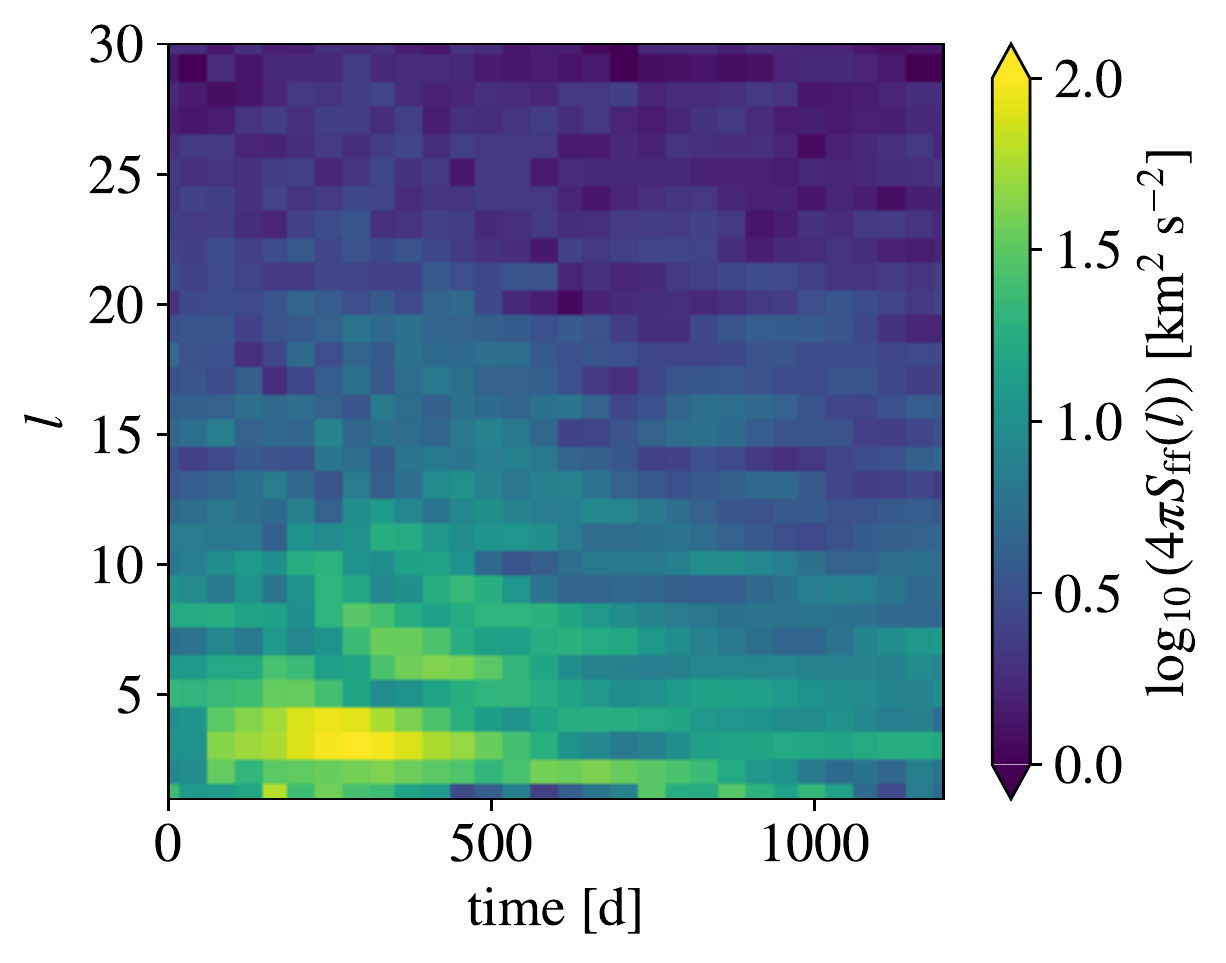}
    \includegraphics[width=\columnwidth]{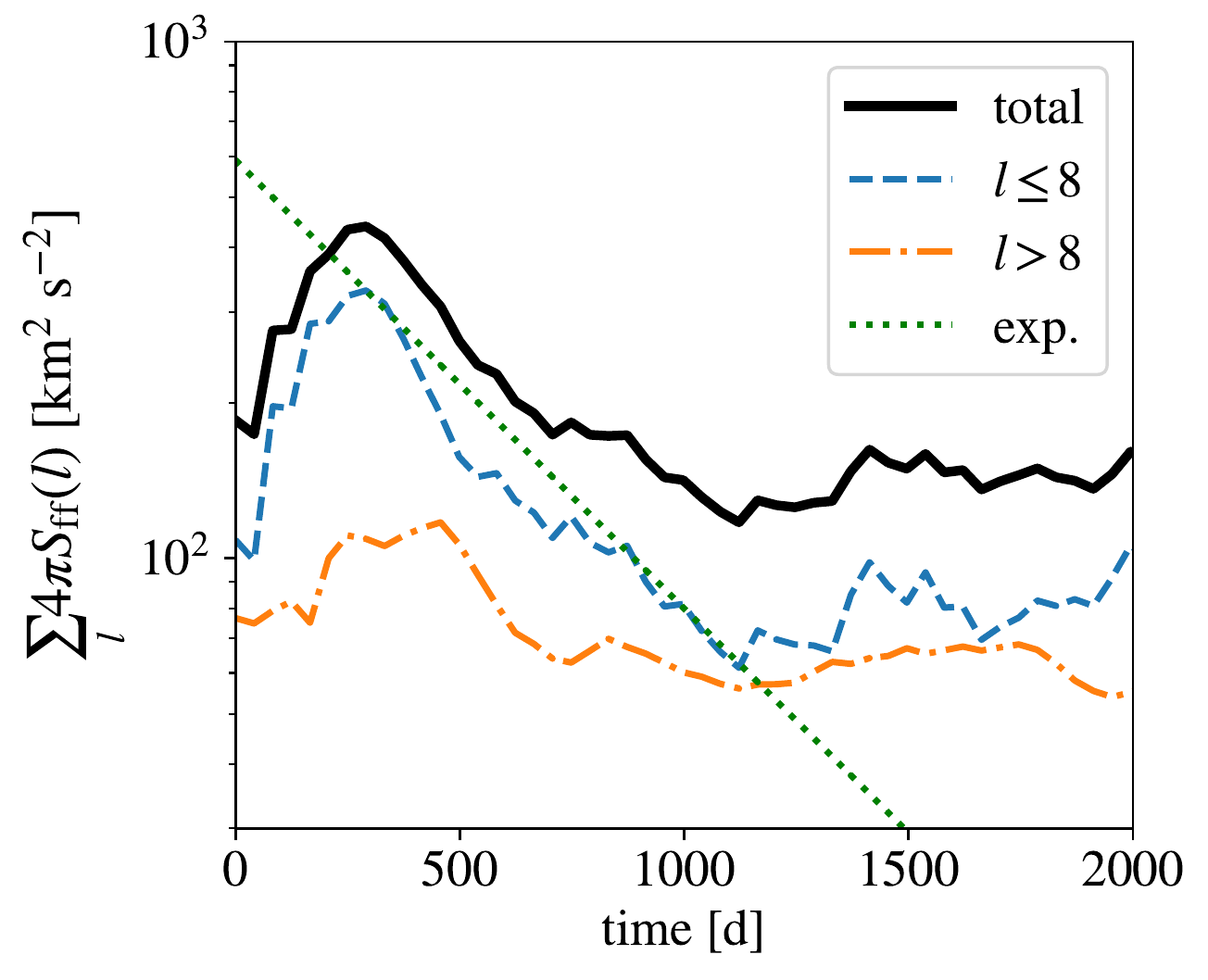}
    \caption{Left: Power spectrum of radial velocities during, and following, the breakout of the convective plume. The initial injection of power into moderate degree modes ($l\sim 3-4$) decays over time in intensity and dissipates to smaller scales through convective turbulence. Right: Integrated power in low-degree ($l\leq8$) and high degree ($l>8$) modes as a function of time. The power in the large-scale, low-degree oscillations decays exponentially in time, as predicted by the theory of convective dissipation. }
    \label{fig:powerspec}
\end{figure*}

In the aftermath of the anomalous convective plume's rise our model star remains disturbed for some time as the plume spreads and dissipates. We analyze this dispersion from the perspective of the flow at the stellar surface.

We compute the time-dependent power spectrum of the surface radial velocity  using {\tt shtools} \citep{2018GGG....19.2574W}. Specifically, the sum
\begin{equation}
    S_{\rm ff}(l) = \sum_{m=-l}^{l} v_{lm}^2,
\end{equation}
is the power per degree, $l$, where
\begin{equation}
    v_{lm} = \frac{1}{4\pi} \int_{\Omega} v_r Y_{lm} d\Omega,
\end{equation}
is an integral of the radial velocity convolved with the spherical harmonics over the surface angular area. The normalization is such that the sum over $l$,
\begin{equation}
    4\pi \sum_l S_{\rm ff}(l) = \int v_r^2 d\Omega ,
\end{equation}
 equals the integrated velocity squared over the surface. 

The left-hand panel of Figure \ref{fig:powerspec} shows the emergence of elevated power in the $l\sim 3-4$ nonradial degrees of the power spectrum at $t\lesssim 400$~d. This reflects the local, rather than global, disturbance of the rising plume. Meanwhile, the same convective motions that drive the upwelling provide a dissipation mechanism. As the plume interacts with its surroundings, it drives a turbulent cascade, elevating oscillatory power at all scales. The cascade leads to the dispersion of energy from a coherent large-scale plume to increasingly smaller scales and, eventually, to heat.  Figure \ref{fig:powerspec}, captures this dissipation in action. While the initial power of the convective plume is deposited in $l\sim 3-4$ modes, we see a flow of power to progressively higher degree modes (with smaller associated size scales on the stellar surface) over the next 100~d. As time progresses from $\sim 200$~d to $\sim 400$~d, a distinct transfer from $l<10$ out to higher $l$ is seen.


The approximate dissipation rate, $\gamma \equiv \dot E / E$, is related to the scale of convective eddies and their characteristic velocities. The origin of this effect is the superposition of the random motion of turbulent convection on the coherent, wavelike motions of a large-scale oscillation. From these scalings \citep{1977A&A....57..383Z,1995A&A...296..709V,2020MNRAS.496.3767V} estimate
\begin{equation}
    \gamma \approx \frac{M_{\rm env}}{M}\left( \frac{L}{MR^2} \right)^{1/3}
\end{equation}
where $M_{\rm env}$ is the convective envelope mass, which is a portion of $M$, the entire stellar mass. A significant point is that this damping rate depends on the properties of the convection, not on the energy of the feature being damped, for example. As a result, exponential damping of oscillatory power is expected. 

For our simulated system, we may substitute $M$, $R$, $L$ to compute the estimated damping rate, where $M_{\rm env}/M=1$ because we model only the fully convective envelope. We find a characteristic damping timescale of $\gamma^{-1} = 560$~d. The right-hand panel of Figure \ref{fig:powerspec} examines the extent to which this is realized in our model. We sum over spherical harmonic degrees to examine the total power and its contribution from low-degrees $l\leq 8$ and higher degrees $l > 8$. A first observation is that a clear peak emerges after the start of the simulation as the plume rises and breaks out of the surface. The low-degrees peak first, followed by a transfer to the higher-degrees as expected from the turbulent cascade. Over the following $\sim 1000$~d, the power at large scales (low degrees) fades. We overplot an exponential decay from the peak with a characteristic timescale of 500~d. We find that this provides an excellent description of the power in the low degrees as they fade by nearly an order of magnitude. It is quite compelling that the timescale is within $\sim 10$\% of that predicted by the order-of-magnitude estimate.

\section{Application to Betelgeuse}

\subsection{Betelgeuse after the Great Dimming}

\begin{figure*}[tbp]
    \centering
    \includegraphics[width=\textwidth]{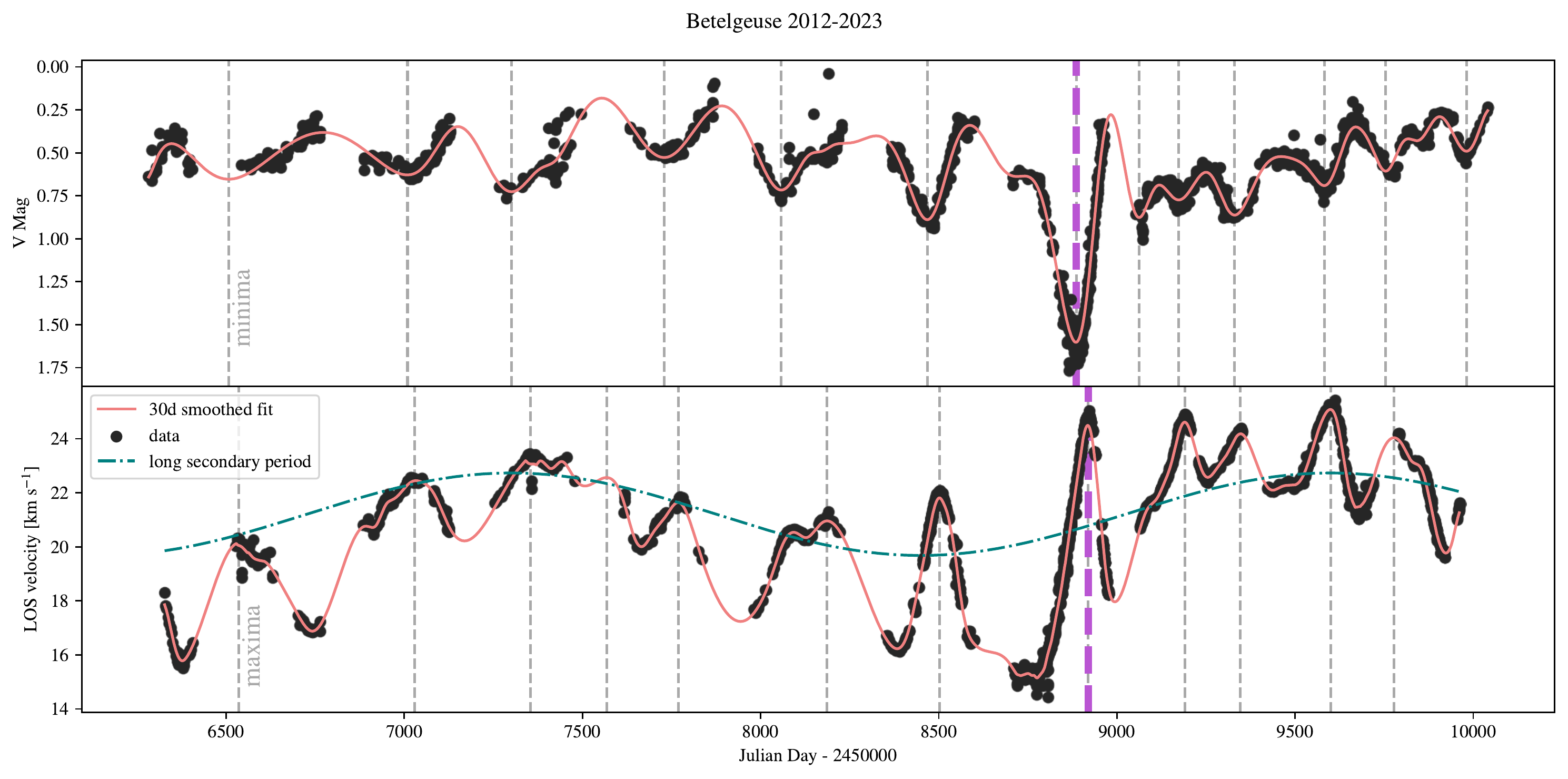}
    \caption{ Betelgeuse's light and line-of-sight radial velocity curves in the interval 2012-2023. The upper panel shows V-band magnitudes tabulated by the AAVSO, with times of minima marked. The lower panel shows radial velocities from STELLA \citep{2022csss.confE.185G}, with times of radial velocity maxima marked. The systemic velocity is 20.57~km~s$^{-1}$. In each case we show a smoothed spline fit to the curves, from which the maxima and minima are identified. In the radial velocity data, we also show a best-fit sine curve for the long secondary period.   }
    \label{fig:lcrv}
\end{figure*}

In the time since its Great Dimming, Betelgeuse's photometric and radial velocity curves have changed significantly. These changes are reported in detail by \citet{2022ApJ...936...18D}, \citet{2022csss.confE.185G}, and \citet{2023NewA...9901962J}. In particular, there is clear evidence for a dramatic change in periodicity before and after the great outburst. 

Before the outburst, the dominant photometric periods were 2170$\pm$270~d, 417$\pm17$~d, 365$\pm75$~d, and 185$\pm4$~d  \citep{2023NewA...9901962J}, where we note that these values are consistent with previous estimates \cite[e.g.][]{2006MNRAS.372.1721K,2020ApJ...902...63J}. The Lomb-Scargle periodogram of the radial velocity data exhibits peaks at frequencies corresponding to 2205~d, 394.5~d, and 216~d \citep{2022csss.confE.185G}. By fitting a sine curve, \citet{2022csss.confE.185G} find a 2169$\pm 5$~d longest period. Indeed this $\sim 6$~yr period was first derived in data approximately 100~yr ago by \citet{1928MNRAS..88..660S}. We note that the photometric 365~d$\approx1$~yr period is likely due to aliasing with Betelgeuse's annual visibility.

The $\sim$395--417~d period corresponds to the star's fundamental mode \citep{2006MNRAS.372.1721K,2020ApJ...902...63J,2022csss.confE.185G}. The 2170~d period is a ``long secondary period," of unknown origin \citep{2006MNRAS.372.1721K,2009MNRAS.399.2063N,2009JRASC.103...11P,2009ApJ...707..573W,2010ApJ...725.1170S,2021ApJ...911L..22S,2022csss.confE.185G}. Comparing to our computed mode frequencies in Table \ref{tab:modes}, we see that the $\sim 200$~d periods are very likely a first overtone of the radial mode, as suggested by \citet{2020ApJ...902...63J} and \citet{2022csss.confE.185G}. Finally, we note that if the peak at 365~d is physical, it could be a low-order nonradial fundamental mode, like the $l=2$ fundamental mode, with a predicted period of $378$~d in our model star.

After the outburst, \citet{2023NewA...9901962J} note that the dominant mean period is 230$\pm 29$~d. We explore this further in Figure \ref{fig:lcrv}, which shows Betelgeuse's V-band light curve and radial velocity curve in the interval 2012-2023. The photometric data is compiled by and retrieved from the AAVSO.\footnote{ \url{https://www.aavso.org/}} The radial velocity data is from STELLA and is reported by \citet{2022csss.confE.185G}. For both datasets, we derive 30-day smoothed-spline fits, which are shown with solid lines rather than points. We identify minima and maxima in the light and radial velocity curves using a 60~d ($\pm 30$~d) window. This effectively filters shorter-timescale local minima and maxima from being considered, and identifies the longer-timescale features. These minima and maxima are shown with dashed vertical lines, with the extremum corresponding to the Great Dimming episode highlighted in purple. Finally, we show a best-fit sine curve to the radial velocity data that corresponds to the long secondary period to provide context for how this feature effects the radial velocities. 

\begin{figure}
    \centering
    \includegraphics[width=\columnwidth]{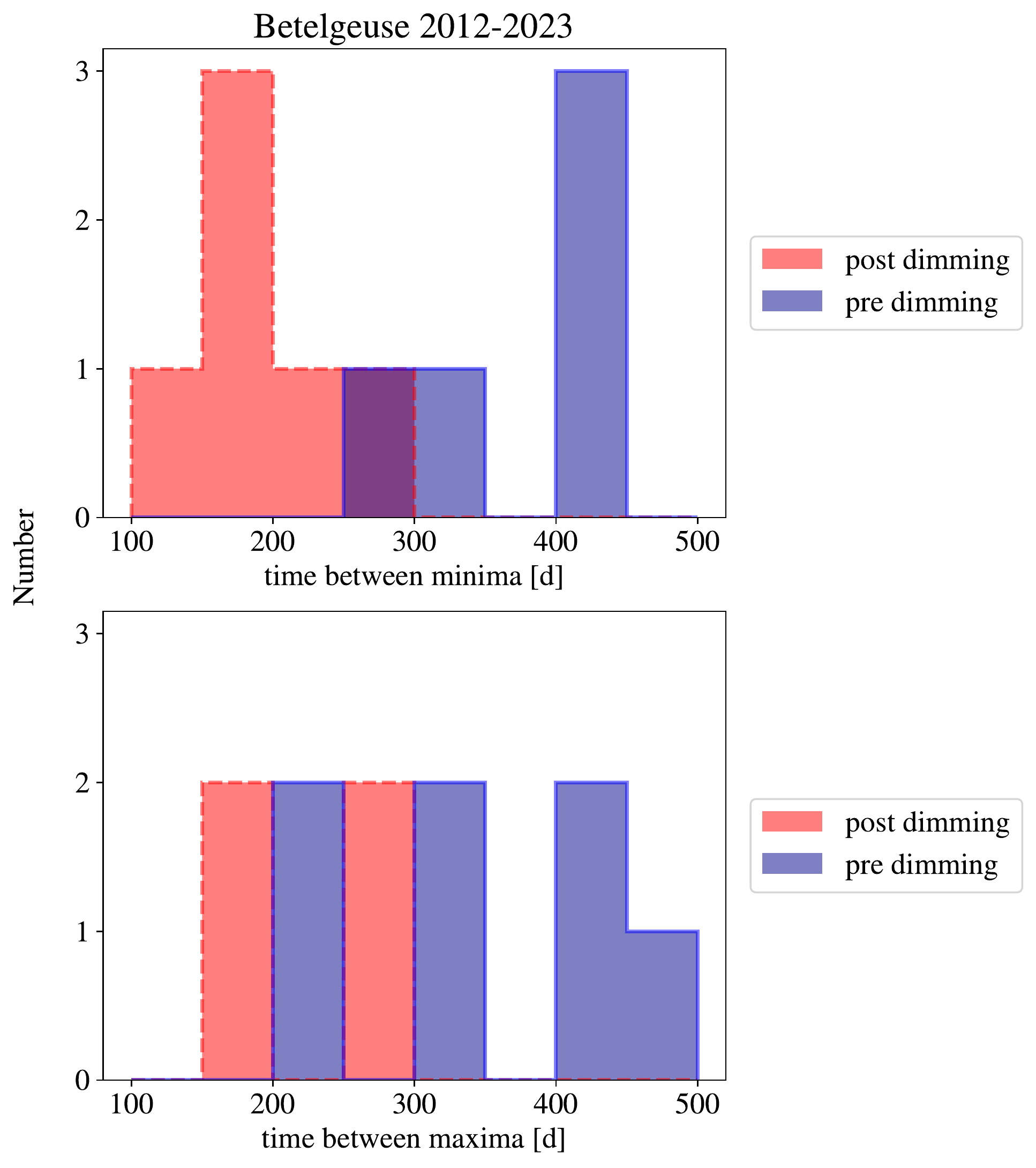}
    \caption{ Histograms of time between successive photometric minima (top panel) or radial velocity maxima (lower panel). Minima and maxima are identified with a 60~d window function that suppresses short-timescale measurement noise and local extrema. } 
    \label{fig:dthist}
\end{figure}

From Figure \ref{fig:lcrv}, we see that similar cycles emerge in the light and radial velocity curves, as hinted by the characteristic frequencies derived in previous work. Photometric minima approximately correspond to radial velocity maxima. However, Figure \ref{fig:lcrv} also illustrates that the variability both before and after the dimming is not strictly periodic. To this end, Figure \ref{fig:dthist} measures the time between successive minima or maxima in the photometry and spectroscopy, respectively. While pre-outburst, the typical separation between cycles is $\sim 300-400$~d, after the outburst, the typical cycle is much shorter, in the range of $\sim 150-270$~d. 

While neither behavior is monoperiodic, we note the near complete distinction between the pre-outburst and post-outburst periodicity. From this analysis it is abundantly clear that the characteristic timescale of Betelgeuse's variability changed dramatically from before to after the Great Dimming. In terms of the characteristic mode frequencies (e.g. Table \ref{tab:modes}) this change likely represents a switch from pulsation primarily in the fundamental mode to pulsation primarily in overtone modes, either of the radial mode or the low-degree nonradial modes.

\subsection{Convective Outburst}

As shown in Figure \ref{fig:lcrv}, just prior to Betelgeuse's dimming, the surface radial velocities remained outflowing for nearly 400~d,  with the LOS velocity remaining below the systemic velocity by several km s$^{-1}$ from (JD-2450000)$\approx 8550-8950$.  
This is much longer than expected within a more typical pulsational cycle, as can be seen by comparison to the more sinusoidal epochs of the remainder of the line of sight radial velocity curve \citep{2022ApJ...936...18D}. Because the length of this plateau is longer than the stellar dynamical time (or any of the typical modes e.g. as tabulated in Table \ref{tab:modes}) we know it must trace the slower,  convective, rather than oscillatory, motion \citep{2022ApJ...936...18D}.

The line-of-sight radial velocity offset from the systemic velocity ($\sim 5$~km~s$^{-1}$) during this velocity plateau is lower than the peak  $\sim 20$~km~s$^{-1}$ local flow velocities seen in our model (Figures \ref{fig:slice} and \ref{fig:vr_ray}). However, the line-of-sight velocity plotted in Figure \ref{fig:lcrv} is averaged over the observable stellar disk, and is therefore more directly comparable to the disk-averaged line of sight velocity computed in Figure \ref{fig:vlos}. In particular, an observer oriented so as to face the breakout of the plume (e.g. the $+x$-direction in the model) sees an extended plateau of negative velocities as material continuously rises.  These features appear to track extremely well with the order of magnitude of the observed radial velocities and their duration.

Thus, because the breakout of the convective plume covers only a fraction of the observable disk of the star, the measured line-of-sight velocities may represent only a portion of the local outflow speeds. 
By comparison, \citet{2021A&A...650L..17K} see evidence for a range of radial velocities up to 10~km~s$^{-1}$ through tomography of the atmosphere in 2019-2020. The presence of (local) flows as fast as $\gtrsim 20$~km~s$^{-1}$ in our model suggests that in small regions, some material may be accelerated to near the escape velocity and lofted to large radii. This material could then contribute to the obscuration hypothesized to cause the dimming itself \citep{2021Natur.594..365M}. The presence of a several-hundred day plateau of outflowing velocity (with disk-averaged magnitude of several km~s$^{-1}$) in both the data and our modeling strongly suggests a rising convective plume is responsible for the immediate pre-dimming features in Betelgeuse.

\subsection{Mode Switching}

Betelgeuse appears to have undergone a sudden switch from its fundamental oscillation mode to a combination of overtones in conjunction with its Great Dimming, as discussed in the context of Figures \ref{fig:lcrv} and \ref{fig:dthist}. Mode switching has been observed in convecting red giants and supergiants. R Doradus underwent several switches in the interval from 1950 to 1970 between a presumed fundamental and overtone modes \citep{1998MNRAS.301.1073B}. RV Andromeda underwent a brief mode switch followed by the superposition of long and shorter period oscillations \citep{1991AJ....101.1043C}. S Aquila and U Bootis were noted to exhibit switches followed quickly (after $<1000$~d) by reversions to the fundamental mode. 

We argue that the surface disturbance of the rising convective plume is crucial in driving the switch to the overtone.  Figures \ref{fig:slice} and \ref{fig:vr_ray} show that the plume rises from the location where it is initially seeded. Material exterior to the rising plume is dramatically disturbed, while material interior to this radius is essentially unmodified initially. Additionally, the plume represents a nonradial disturbance, and covers only a fraction of the stellar surface. 

While our model convective star is static, Betelgeuse is oscillating in the radial fundamental mode. In the case of an ongoing oscillation, portions of the star not disturbed by the spurious plume continue their oscillatory behavior. By comparison to the line-of-sight velocity curves in Figure \ref{fig:lcrv}, we suggest that the phase of this oscillation is significant. The undisturbed portions of the stellar surface and interior progress through maximum expansion around (JD-2450000)$\approx 8600$, then oscillate toward contraction. Meanwhile, the expanding plume's radial velocity is locally much larger than the several km~s$^{-1}$ oscillatory velocity. Though the bulk of the star may be oscillating inward at (JD-2450000)$\approx 8600$--8800, the surface layers in the vicinity of the plume continue to rise, creating the plateau feature of outflow. 

If we were able to look inside the star at (JD-2450000)$\approx 8800$, we would see that the inner portion of the stellar envelope moves inward while the outer portion still moves outward. This new envelope configuration matches more closely that of the first overtone, because there is now a radial node in the velocity profile (see the mode eigenfunctions of Figure \ref{fig:ef}). As the plume dissipates, the oscillatory phase offset between the surface and interior remains. The surface moves outward while the interior moves inward. The period of this overtone oscillation is shorter than that of the fundamental mode, leading to the decrease in cycle spacing seen in the aftermath of the outburst and dimming.

\subsection{Future Reversion to Fundamental Mode?}
Red supergiants tend to pulsate primarily in the fundamental mode \citet{2006MNRAS.372.1721K}. We hypothesize that this relates to the strength of their convective flows in relation to their oscillation amplitudes. In our model of Betelgeuse, interior radial velocities are on the order of $10-20$~km~s$^{-1}$ (e.g. as seen in Figure \ref{fig:slice}), comparable to the $\sim 9$~km~s$^{-1}$ photospheric dispersion noted by \citet{2008AJ....135.1450G}. Comparatively a fundamental mode with surface amplitude of $\sim 2$~km~s$^{-1}$ has interior bulk velocity of $\lesssim1$~km~s$^{-1}$ at half the stellar radius (see the mode eigenfunctions in Figure \ref{fig:ef}). The relative magnitude of these velocities means that the local flow of convection can easily surpass the bulk oscillatory motion. This scenario can be likened to a situation in which the ``signal" -- the oscillation -- has smaller amplitude than the random ``noise" of the turbulent convection. 

For an overtone oscillation to persist over the long term, a radial node in the velocity field would need to survive despite being intercepted by convective plumes with larger scale height (compare the mode eigenfunctions of Figure \ref{fig:ef} to the large-scale convective plumes in Figure \ref{fig:slice}). This is sustainable in stars with less vigorous convection, where the convective velocities are less than the oscillatory velocity \citep[e.g. as is the case for convection in Cephieds, as described by][]{2014A&A...563L...4N}. In red supergiants like Betelgeuse, we suggest that overtones are not generally long-lasting because they are erased by large-scale, powerful convective flows.   Indeed, this conclusion aligns with the observed mode reversals following switches to overtones in R Dor, RV And, S Aql, and U Boo \citep{1991AJ....101.1043C,1998MNRAS.301.1073B}.

When applied to Betelgeuse, this suggests that over time, the post-outburst overtone ringing that we observe may fade, or switch back to the fundamental mode.  As discussed in relation to the dissipation of the initial outburst, the damping time relates to the properties of the convection. Scaled to Betelgeuse's parameters, the characteristic damping time of the convection is 
\begin{eqnarray}
    \gamma^{-1} \approx 10^3~{\rm d} &\left( \frac{M_{\rm env}/M }{0.7} \right)^{-1} \left( \frac{L}{10^5 L_\odot} \right)^{-1/3}  \nonumber  \\
    \times &\left( \frac{M}{17.6 M_\odot} \right)^{1/3} \left( \frac{R}{770 R_\odot} \right)^{2/3} . 
\end{eqnarray}
Thus each $\sim 10^3$~d, we might expect an e-fold decrease in the power in Betelgeuse's overtone oscillations. The power is proportional to the amplitude squared, so each e-fold decrease in the power comes with a $\sqrt{e}$ decrease in amplitude. Thus, the amplitude e-folding time is $2\gamma^{-1}$ or $\sim 2\times10^3$~d$\approx 5.5$~yr. 

\section{Conclusions}

This paper has described a model for the Great Dimming of Betelgeuse that connects its photometric and spectroscopic behavior to the vigorous convection in its envelope. We draw the following key conclusions:
\begin{enumerate}
    \item Particularly hot regions can arise from rare, but strong, collisions in the turbulent stellar interior. These over-pressure regions buoyantly rise and spread, driving a shock ahead of them as they break out of the stellar surface (Figure \ref{fig:slice}).
    \item Convection dissipates and damps the perturbation excited by the breakout of the plume through the turbulent cascade  with a characteristic timescale that can be estimated from the bulk properties of the star (Figure \ref{fig:powerspec}). 
    \item The breakout of such an anomalous convective plume can explain the long-term $\sim400$~d outflow observed just prior to Betelgeuse's dimming episode of 2020 (Figures \ref{fig:vlos} and \ref{fig:lcrv}).  
    \item In the case of Betelgeuse, the pre-existing fundamental mode oscillation was perturbed into a combination of overtones (Figures \ref{fig:lcrv} and \ref{fig:dthist}). This may occur as the surface and interior lose phase coherence while the convective plume rises and breaks out. 
    \item We predict that convective damping will act to reduce the overtone oscillation of Betelgeuse on a characteristic timescale $\sim 2000$~d, with a return to the fundamental mode oscillation possible after that time. 
\end{enumerate}
As a close, well-studied red supergiant, Betelgeuse reveals how enigmatic these objects remain \citep[e.g.][]{2017ars..book.....L}. In particular, it appears possible that the turbulent cascade driven by convection can explain Betelgeuse's recent outburst and mode switching. It hints at a connection between convective perturbations \citep{2006MNRAS.372.1721K}, mode switching \citep{1991AJ....101.1043C}, and episodes of enhanced mass loss in red supergiants \citep{2019A&A...623A.158H}. The detailed study of Betelgeuse may serve as a prototype for  related events, such as the recently observed extreme dimming of an extragalactic red supergiant \citep{2022ApJ...930...81J}.  

As we await Betelgeuse's future evolution, our modeling suggests that its subsequent oscillatory behavior can offer many insights into the nature of its dramatically-convective envelope and how massive stars pass through the giant-branch phases of stellar evolution.  Our hydrodynamic model's approximate treatment of radiative cooling at the stellar surface does not allow us to track a photosphere location or its temperature accurately, but there are important observational constraints on these properties \citep[e.g.][]{2020ApJ...899...68D,2022ApJ...936...18D}. This suggests that there is much to learn from future, more sophisticated modeling.

\begin{acknowledgements}
We thank N. Evans and M. Vick for helpful discussions related to this work.  M.M. gratefully acknowledges support from a Clay Postdoctoral Fellowship. This work used the Extreme
Science and Engineering Discovery Environment (XSEDE), which
is supported by National Science Foundation grant No. ACI1548562. In particular, use of XSEDE resource Stampede2 at TACC through allocation TG-AST200014 enabled this work. This research is supported in part by HST grants HST-GO-15641 and HST-GO-16655 to the Smithsonian Astrophysical Observatory. A.A. gratefully acknowledges support from the National Science Foundation Graduate Research Fellowship under Grant No. DGE 1752814 and the Gordon and Betty Moore Foundation through Grant GBMF5076. A.L. was supported in part by Harvard's Black Hole Initiative, which is funded by GBMF and JTF grants. 
\end{acknowledgements}

\software{IPython \citep{PER-GRA:2007}; SciPy \citep{2020SciPy-NMeth};  NumPy \citep{van2011numpy};  matplotlib \citep{Hunter:2007}; Astropy \citep{2013A&A...558A..33A}; MESA \citep{2011ApJS..192....3P,2013ApJS..208....4P,2015ApJS..220...15P,2018ApJS..234...34P,2019ApJS..243...10P}; GYRE \citep{2013MNRAS.435.3406T,2018MNRAS.475..879T,2020ApJ...899..116G}; Athena++ \citep{2020ApJS..249....4S}.  }

\bibliographystyle{aasjournal}

\appendix

\section{Mode Frequencies}

The dimensionless and dimensional characteristic frequencies of various degree, $l$, and radial order, $n_{\rm pg}$ are listed in Table \ref{tab:modes}. We note that the dimensionless frequencies, in units of  $(GM/R^3)^{1/2}$, are only mildly sensitive to the evolutionary state of the model within the core-helium burning phase. This means that the dimensional mode phases can be scaled to different $M$ and $R$ accordingly. The growth rate, $\eta$ is dimensionless in the range $-1\leq \eta \leq 1$. Values of $\eta >0$ indicate energy generation over a mode cycle and that a particular mode is unstable and will grow. Values of $\eta <0$ indicate damped modes. 

\begin{figure}
    \centering
    \includegraphics[width=\textwidth]{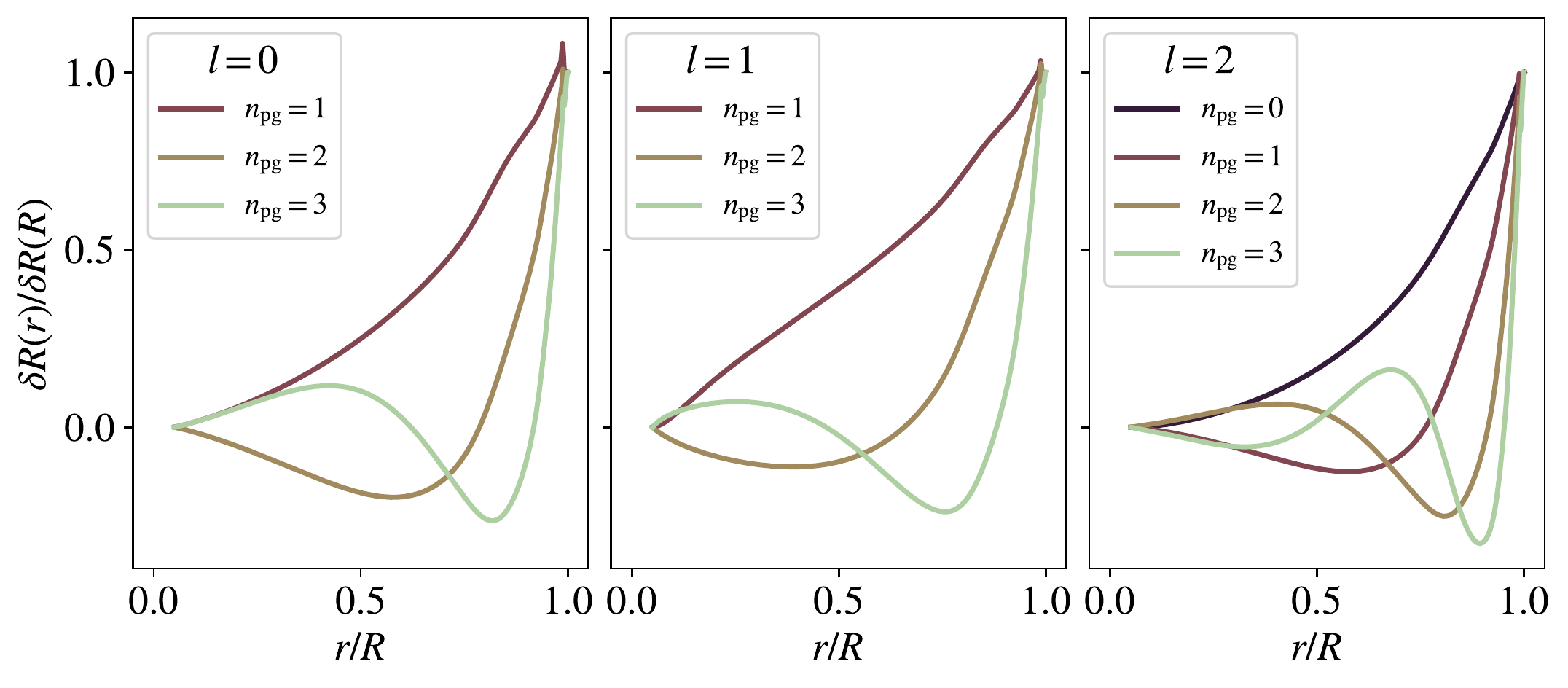}
    \caption{Eigenfunctions of radial $(l=0)$ and nonradial $(l\geq1)$ modes, classified by $n_{\rm pg}$ the fundamental mode's eigenfunction is monotonic, while the overtones exhibit increasing numbers of radial nodes.  }
    \label{fig:ef}
\end{figure}

Figure \ref{fig:ef} shows eigenfunctions of the radial displacement of $l=0$, 1, and 2 modes, scaled to the radial displacement at the surface. All of these modes have their largest amplitude at the surface, and the higher radial order harmonics have increasing numbers of inflection points.

\begin{table}
    \centering
\begin{tabular}{cccccc}
\hline
$l$ & $n_p$ & $n_{pg}$ & $\omega$ & $P$ & $\eta$ \\
- & -  & - & [$(GM/R^3)^{1/2}$] & [d] & - \\
\hline 
0 & 1 & 1 & 1.457 & 405.7 & 0.36 \\
0 & 2 & 2 & 3.104 & 190.4 & 0.53 \\
0 & 3 & 3 & 4.915 & 120.2 & 0.74 \\
\hline
1 & 0 & 1 & 1.091 & 541.6 & 0.4 \\
1 & 1 & 2 & 2.468 & 239.4 & 0.47 \\
1 & 2 & 3 & 4.225 & 139.9 & 0.66 \\
\hline
2 & 0 & 0 & 1.561 & 378.6 & 0.35 \\
2 & 1 & 1 & 3.02 & 195.7 & 0.49 \\
2 & 2 & 2 & 4.827 & 122.4 & 0.72 \\
2 & 3 & 3 & 6.457 & 91.5 & 0.79 \\
\hline
3 & 0 & 0 & 1.863 & 317.3 & -0.04 \\
3 & 1 & 1 & 3.451 & 171.2 & 0.51 \\
3 & 2 & 2 & 5.286 & 111.8 & 0.75 \\
3 & 3 & 3 & 6.992 & 84.5 & 0.77 \\
\hline
4 & 0 & 0 & 2.11 & 280.0 & -0.32 \\
4 & 1 & 1 & 3.829 & 154.3 & 0.52 \\
4 & 2 & 2 & 5.675 & 104.1 & 0.77 \\
4 & 3 & 3 & 7.501 & 78.8 & 0.76 \\
\hline 
\hline
\end{tabular}
    \caption{Properties of oscillatory modes in our MESA model of Betelgeuse as computed by GYRE \citep{2013MNRAS.435.3406T,2018MNRAS.475..879T,2020ApJ...899..116G}. Modes are listed by their spherical harmonic degree $l$, with $l=0$ representing radial modes and $l>0$ representing non radial modes. The integer $n_p$ is the p-mode classification, and $n_{pg}$ is the $p$ and $g$ mode classification. For $l=0$ and $l=1$, $n_{\rm pg}=1$ is the fundamental mode, while for $l\geq 2$, $n_{pg}=0$ is the fundamental mode. Higher $n_{pg}$ represent overtone $p$ modes with radial nodes in their eigenfunctions. Mode frequencies are listed in their dimensionless forms $\omega$,  and scale weakly as the star evolves in the range of interest. Their dimensional conversion to period $P$ in days sensitively traces the mass and radius of the star. Finally, $\eta$ is the growth rate of the mode, defined such that $-1<\eta < 1$, where unstable (growing) modes have $\eta>0$ and stable (damped) modes have $\eta<1$ \citep{2018MNRAS.475..879T}.  }
    \label{tab:modes}
\end{table}

\section{Numerical Tests}

\begin{figure}
    \centering
    \includegraphics[width=0.49\columnwidth]{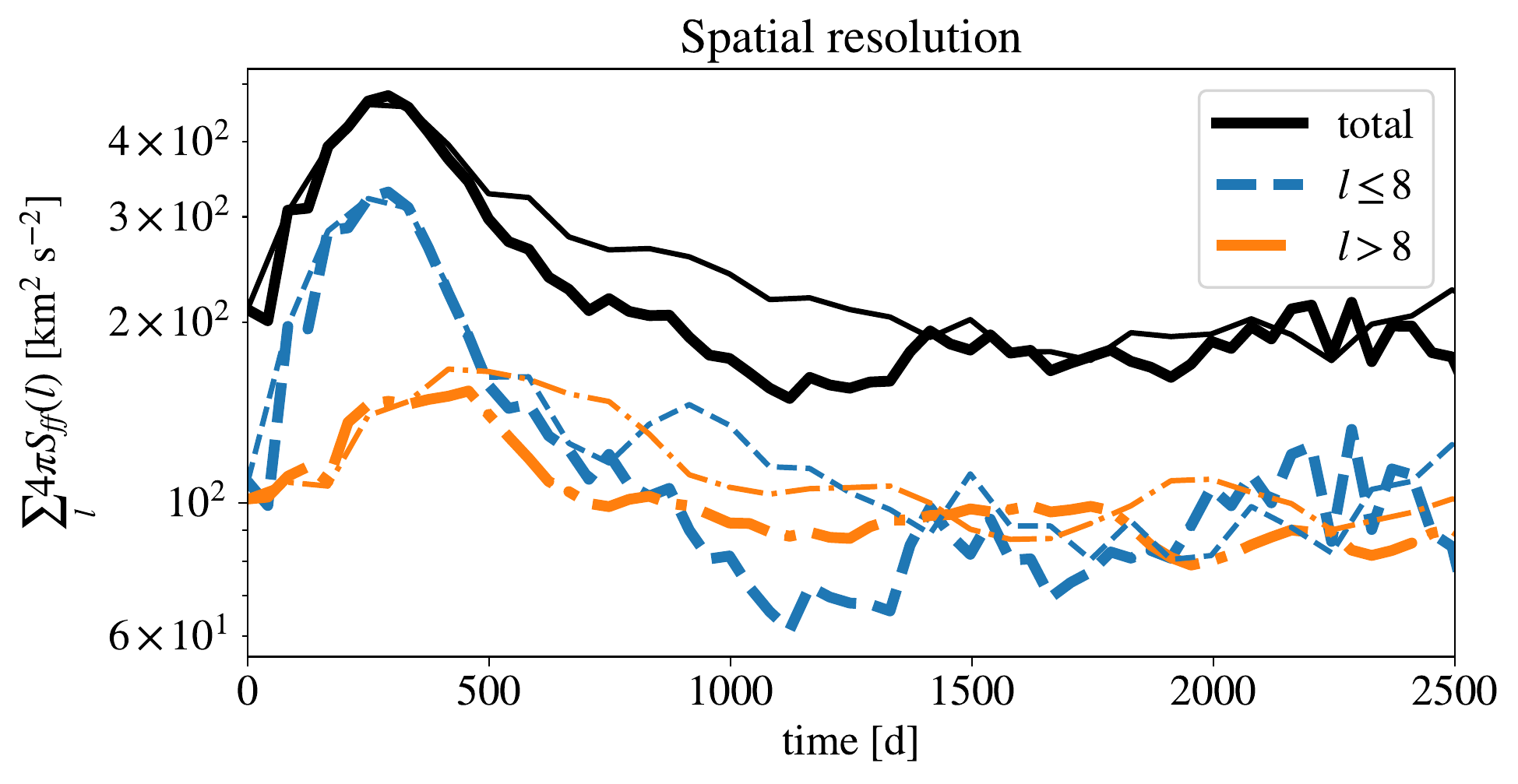}
    \includegraphics[width=0.49\columnwidth]{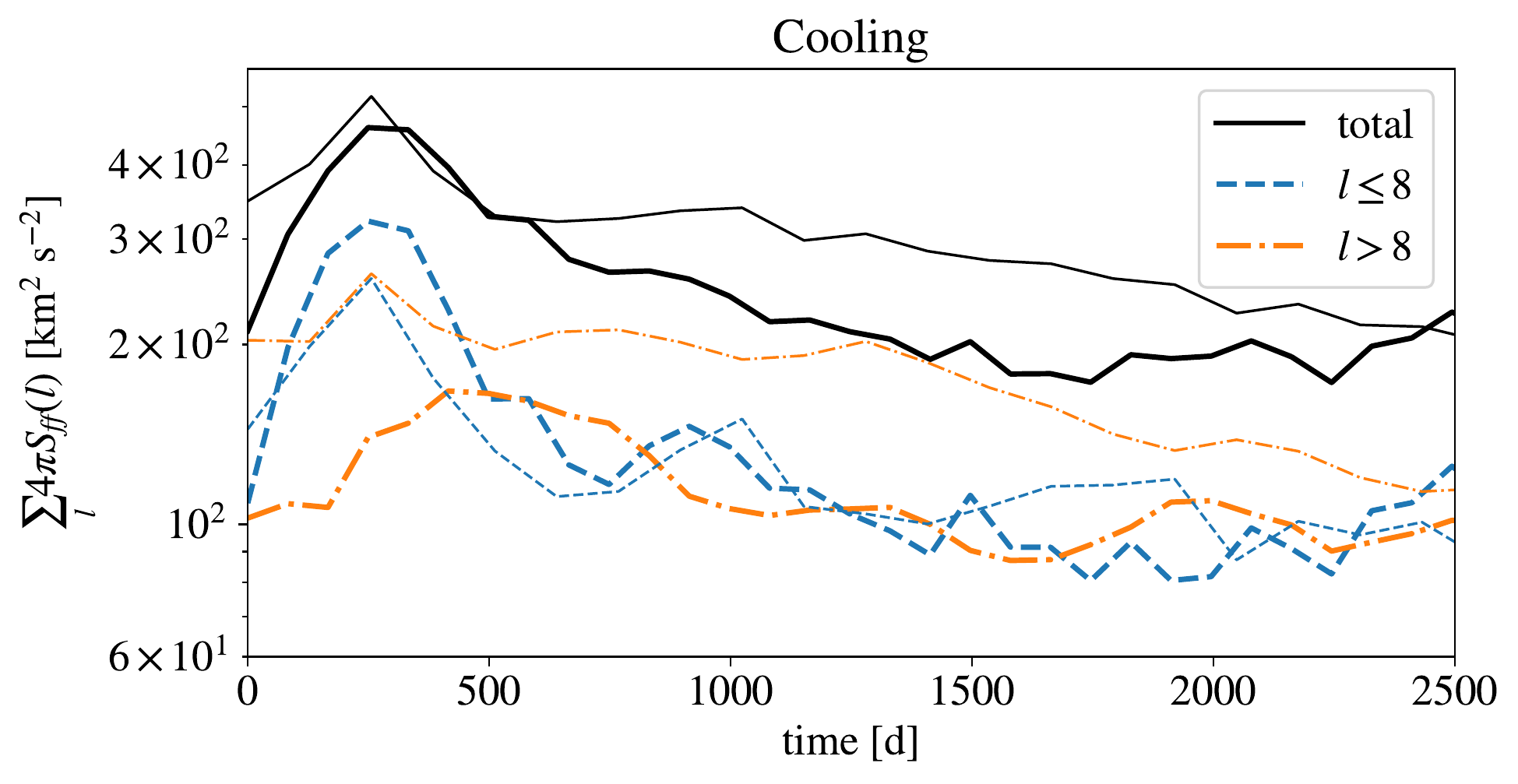}
    \caption{Numerical tests. {\it Left Panel:} Comparison of identical initial conditions run at two resolutions. Our fiducial, higher, resolution model has nested meshes each with $256^3$ zones and is shown with thicker lines, while a factor of two lower resolution case (nested meshes of $128^3$ zones) is shown with thinner lines. We find good agreement with the behavior of the models near peak, but find that the power in the lower-resolution model plateaus at slightly higher levels. This indicates to us that the decay rate of the $l\leq 8$ power for the first $\sim 800$~d is not sensitive to the spatial resolution.  {\it Right Panel:} Comparison between two $128^3$ resolution runs with differing cooling. The thicker lines in this panel are the same match the left panel. The thinner lines are a model with more-rapid cooling, evaluated at it's own effective radius of $6r_0$ in code units. We note that the small-scale power $(l>8)$ is higher when the cooling is more efficient. However, the large-scale modes ($l\leq 8$) are largely insensitive to the cooling approximation, just as found in terms of spatial resolution. }
    \label{fig:numerical}
\end{figure}

We tested the sensitivity of our results to numerical resolution and to the properties of the surface cooling. A comparison of the power as a function of time in two resolutions is shown in Figure \ref{fig:numerical}. In our fiducial case, nested layers are resolved with $256^3$ zones. We compare to a model with half the spatial resolution but the same grid structure, such that each nested layer is resolved by $128^3$ zones.  We observe that the large-scale ($l\leq 8$) power is nearly identical in the two resolution cases for the first $\sim 800$~d, after which some differences exist, as might be expected given the chaotic nature of the turbulent flow. The small scale modes, $l>8$, decay slightly more slowly in the lower-resolution model. However, we find that the basic property of the early exponential decay rate in the first $\sim 800$~d is not affected by the spatial resolution. 

Secondly we assess the importance of our approximate cooling scheme at the stellar surface. As described in Section 2.3, in our fiducial runs, we subject surface material to a characteristic cooling time of 10 times the local dynamical time. For comparison, we adopt a choice of $10^{-2}$ of the local dynamical time to ensure that our dynamics are not determined by the properties of our artificial photosphere. Because cooling affects the stellar pressure and density profiels, the scaling of the hyrodynamic model to the MESA simulation differs somewhat. In this case, we rescale such that the stellar radius of $770R_\odot$ is equal to $6r_0$ in code units. The right-hand panel of Figure \ref{fig:numerical} compares the $128^3$-base resolution fiducial case to a model with $128^3$-base resolution and the more-efficient cooling. We note that the small-scale power ($l>8$) is significantly higher in the case of more-rapid cooling. 
However, we find that the most features of importance to our study are not dramatically affected by the cooling parameterization. In particular, the overall power in the low-degree modes and total (near peak) are similar, despite the need to rescale to different radii. Secondly, and most importantly, the decay of the low-degree ($l\leq 8$) power decays similarly in time. 

These tests demonstrate which features of our hydrodynamic models are robust with respect to the numerical choices and parameterizations required. In particular, we find that the power in low spherical harmonic degrees ($l\leq8$) and its, approximately exponential temporal decay are robust with respect to both spatial resolution and the cooling approximation.

\section{Software and Data}

 The data and software needed to reproduce the figures and analysis presented in this paper are available online at \url{https://github.com/morganemacleod/BetelgeuseConvection}. 

\end{document}